\documentclass[pra,amsmath,amssymb, two column]{revtex4}
\usepackage{mathrsfs}
\usepackage{amssymb}

\input epsf.tex
\usepackage{graphicx}
\usepackage{amsthm}

\begin{document}

\title{Detection efficiency and noise in semi-device independent randomness extraction protocol}
\author{Hong-Wei Li$^{1,2,3,4}$,Zhen-Qiang Yin$^{1,4,a}$, Marcin Paw{\l}owski$^{5,b}$, Guang-Can Guo$^{1,4}$ and Zheng-Fu Han$^{1,4,c}$}

 \affiliation
 {$^1$ Key Laboratory of Quantum Information, University of Science and Technology of China, Hefei, 230026,
 China\\
 $^2$State Key Laboratory of Networking and Switching Technology, Beijing University of Posts and Telecommunications, Beijing, 100876, China\\
 $^3$ Zhengzhou Information Science and Technology Institute, Zhengzhou, 450004,
 China\\
 $^4$ Synergetic Innovation Center of Quantum Information and Quantum
Physics, University of Science and Technology of China, Hefei
230026, China \\
 $^5$Institute of Theoretical Physics and
Astrophysics, University of Gdansk, 80-952 Gdansk, Poland\\
 }

 \date{\today}
\begin{abstract}

In this paper, we analyze several critical issues in semi-device
independent quantum information processing protocol. In practical
experimental realization randomness generation in that scenario is possible only if the efficiency of the detectors used is above a certain threshold. Our analysis shows that the
critical detection efficiency is $\frac{\sqrt{2}}{2}$ in the
symmetric setup, while in the asymmetric setup if one of the bases has perfect critical detection
efficiency then the other one can be arbitrarily close to $0$. We also analyze the semi-device independent
random number generation efficiency based on different averages of guessing
probability. To generate more randomness, the proper
averaging method should be applied. Its choice depends on the value of a certain dimension witness. More importantly, the general analytical
relationship between the maximal average guessing probability and
dimension witness is given.

\end{abstract}
\maketitle

{\it Introduction -} A bound on the Hilbert space dimension is an
important resource for quantum information processing, which can
increase the performance of the quantum key distribution (QKD) and
quantum random number generation (QRNG) protocols to avoid the
attacks exploiting imperfections of the devices. Based on the
certified system dimension, the notion of semi-device independent
(SDI) protocol can be defined in the prepare and measure scenario,
which assumes the knowledge of the dimension of the underlying
physical system but otherwise nothing about the actual physical
implementation of the state preparation and measurement. The first
SDI quantum key distribution (SDI-QKD) protocol was proposed by
Paw{\l}owski and Brunner \cite{marcin}. Then the SDI random number
generation (SDI-RNG) protocol has been proposed, analyzed
\cite{li1,li2,li3,li4,att} and, eventually experimentally realized
\cite{expdw}. All of them use a dimension witness \cite{dw} to
certify randomness of the measurement outcomes.

To guarantee the randomness without relying on assumption on the
internal functioning of the state preparation and measurement
devices, device independent (DI) protocols based on Bell
inequalities were previously proposed \cite{di1,di2,di3}. However,
the DI protocols require high detection efficiency to avoid the, so
called, detection loophole \cite{Pearle}. The critical detection
efficiency of the maximally entangled state to exclude the
possibility of a local hidden variable description is $82.8\%$
\cite{mermin}, when Alice and Bob have measurement setups with equal
detection efficiencies. This requirement can be reduced to
$\frac{2}{3}$ if the non-maximally entangled state are used
\cite{eberhard}. Similar to the detection loophole in the DI case,
SDI protocol also require the measurement setup in Bob's side to
have high detection efficiency. Now, there are two methods to solve
this problem, both based on making some additional assumption. In
\cite{att,brunner1,brunner2} a nonlinear dimension witness to
certify generated random numbers is used, however the state
preparation device and measurement device are assumed to be
independent \cite{yin1,yin2,li5,li6}. More recently, Canas et al.
applied a trusted blocking device to solve this problem
\cite{marcin2}.

In this work we analyze the critical detection efficiency without any additional
assumption. We prove that the critical detection efficiency is
$\frac{\sqrt{2}}{2}$ in the symmetric case (where Bob's two
measurement bases have the same detection efficiency), while it can be arbitrarily close to $0$ in the
asymmetric case (where one of Bob's two measurement bases has
perfect detection efficiency). We also calculate the amount of certified randomness based on different averages of guessing
probability. The result demonstrates that different averaging methods should be applied depending on the dimension
witness values, and that true randomness can be generated if the
dimension witness value is larger than the classical dimension witness
upper bound. More importantly, the analytical relationship between
the dimension witness and maximal average guessing probability is given, which can be directly applied in the future SDI quantum
information protocol research.

{\it Critical detection efficiency in SDI protocol -} A SDI protocol involves two parties: the sender (Alice) and the receiver (Bob). They both get classical input $x$ for Alice and $y$ for Bob. Then Alice sends a state $\rho_x$ to Bob. We do not know what this state is but assume an upper bound on its Hilbert space dimension $d$. Bob chooses a measurement based on $y$ and obtains the outcome $b$. Then the parties estimate the conditional probability distribution $p(b|x,y)$. A dimension witness is a function of this probability distribution. Dimension witness' quantum bound $Q_d$ is the largest value of this function possible to obtain with the communications of systems of dimension $d$. Similarly the classical bound $C_d$ is the largest value possible to obtain with the communication of $\log d$ classical bits.

The powerhorse of SDI protocols has been a task known as $2\to 1$ Random Access Code \cite{RAC}. The objective of the parties is for the sender to encode two classical bits $a_0$ and $a_1$ into a single qubit of communication aiming to maximize the probability of successfully decoding a single bit of the receiver's choice. Both first: QKD \cite{marcin} and QRNG \cite{li1} protocols have been based this code. Protocols from \cite{li2} were based on its generalization. Moreover, the ones using nonlinear witnesses \cite{brunner1,att,marcin2} are also the realizations of the same task but with a different measures of its efficiency. The sets of preparations and measurements which are optimal are the same in all these protocols. Therefore, we will limit our analysis to the simplest case - the one form \cite{li1}.

We take $d=2$ and apply the following dimension witness
 to distinguish between the classical and quantum systems
\begin{equation}
\begin{array}{lll}
T\equiv E_{000}+E_{001}+E_{010}-E_{011}\\
~~~~~~-E_{100}+E_{101}-E_{110}-E_{111},
\end{array}
\end{equation}
where $E_{a_0a_1y}=p(b=0|a_0,a_1,y)=tr(\rho_{a_0a_1}M_y^{b=0})$,
$a_0,a_1,y,b\in\{0,1\}$, $M_y^{b=0}$ is the measurement operator
acting on the two dimensional state $\rho_{a_0a_1}$ with the input
parameter $y$ and the measurement output $b=0$. In the two dimensional
space, the upper bound of $T$ for the classical system is $2$, while
the quantum allow for $T$ up to $2\sqrt{2}$.

In a practical experimental realization, the quantum state maybe undetected by
the receiver due to the quantum channel loss and/or imperfect detector
efficiency. We assume that Bob's two different measurement
bases $\{M_0^{0},M_0^{1}\}$ and $\{M_1^{0},M_1^{1}\}$ have the
detection efficiency efficiency $\eta_0$ and $\eta_1$ respectively.
Similar to the detection loophole analysis in a DI protocol,  Bob will
output 1 when no detectors click, resulting in a modified
conditional probability $\widetilde{E_{a_0a_10}}$ and
$\widetilde{E_{a_0a_11}}$ with finite detection efficiency $\eta_0$
and $\eta_1$. It can be given by

\begin{equation}
\begin{array}{lll}
\widetilde{E_{a_0a_10}}=\eta_0E_{a_0a_10},~~\widetilde{E_{a_0a_11}}=\eta_1E_{a_0a_11}.\\
\end{array}
\end{equation}
 By using the modified
conditional probability $\widetilde{E_{a_0a_10}}$ and
$\widetilde{E_{a_0a_11}}$, the new dimension witness value is
\begin{equation}
\begin{array}{lll}
\widetilde{T}\equiv
\widetilde{E_{000}}+\widetilde{E_{001}}+\widetilde{E_{010}}-\widetilde{E_{011}}\\
~~~~~~-\widetilde{E_{100}}+\widetilde{E_{101}}-\widetilde{E_{110}}-\widetilde{E_{111}}\\
~~=\eta_0E_{000}+\eta_1E_{001}+\eta_0E_{010}-\eta_1E_{011}\\
~~~~-\eta_0E_{100}+\eta_1E_{101}-\eta_0E_{110}-\eta_1E_{111}.
\end{array}
\end{equation}
Since Bob has two measurement bases setup, it is natural to consider symmetric and asymmetric cases, i.e. with $\eta_1=\eta_0$ and $\eta_1\neq\eta_0$
respectively. In the symmetric case, Bob's two measurement bases
$M_0^{b}$ and $M_1^{b}$ have the same detection efficiency
$\eta_0=\eta_1=\eta$, thus the new dimension witness is
\begin{equation}
\begin{array}{lll}
\widetilde{T_1}
=\eta(E_{000}+E_{001}+E_{010}-E_{011}\\
~~~~~~~~~~-E_{100}+E_{101}-E_{110}-E_{111}).
\end{array}
\end{equation}
By applying the quantum dimension witness upper bound $2\sqrt{2}$,
we can get the new dimension witness value $2\sqrt{2}\eta$. To
violate the classical dimension witness upper bound (that is to
guarantee $\widetilde{T_1}>2$), the corresponding critical detection
efficiency in the symmetric case is $\frac{\sqrt{2}}{2}$.

In the asymmetric case, Bob's two measurement bases
$\{M_0^{0},M_0^{1}\}$ and $\{M_1^{0},M_1^{1}\}$ have different
detection efficiency ($\eta_0\neq\eta_1$). This scenario can be
realized by the neutral kaons system \cite{kaon}, where the first
basis can be performed by lifetime measurement quite efficiently,
but the second basis can be performed by strangeness measurement
with small efficiency. Here, we simply assume the first base has the
perfect detection efficiency ($\eta_0=1$), thus the dimension
witness will be transformed to
\begin{equation}
\begin{array}{lll}
\widetilde{T_2}
=E_{000}+\eta_1E_{001}+E_{010}-\eta_1E_{011}\\
~~~~~~-E_{100}+\eta_1E_{101}-E_{110}-\eta_1E_{111}.
\end{array}
\end{equation}

Applying the Levenberg-Marquardt algorithm optimization numerical
calculation method \cite{optimize}, we calculate the maximal
dimension witness value with different detection efficiency $\eta_1$

\begin{equation}
\begin{array}{lll}
maximize: \widetilde{T_2}(\eta_1),\\
 subject~~to:  E_{a_0a_1y}=tr(\rho_{a_0,a_1}M_y^{b=0}),
\end{array}
\end{equation}
where $\rho_{a_0,a_1}=\displaystyle
\frac{1}{2}(I+\vec{s_{a_0a_1}}\cdot\vec{\sigma})$ is the arbitrary
two dimensional quantum state preparation, $\vec{s_{a_0a_1}}$ is the
Bloch vector and $\vec{\sigma}$ is the Pauli matrix vector.
$M_y^{b=0}=c_yI+\vec{t_y}\cdot\vec{\sigma}$ is the arbitrary two
dimensional positive-operator valued measure (POVM) ( $\vec{t_y}$ is
the Bloch vector), which should satisfy the semi-definite
restriction ( $M_y^{0}$ and $M_y^{1}$ respectively have two nonzero
eigenvalues, $M_y^{0}+M_y^{1}=I$). The corresponding calculation
result is given in Fig. 1.

\begin{figure}[!h]\center
\resizebox{9cm}{!}{
\includegraphics{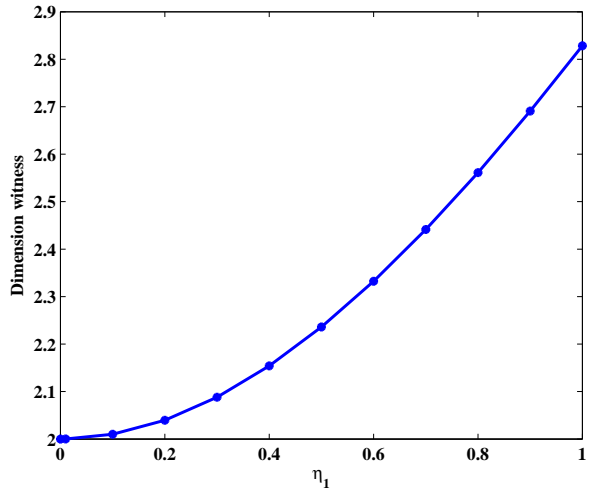}}
\caption{Maximal dimension witness as a function of detection
efficiency in the asymmetric case.}
\end{figure}
From the calculation result, we can find that the quantum dimension
witness value will violate the classical dimension witness upper
bound with arbitrary nonzero detection efficiency $\eta_1$, which
obviously improves the previous critical detection efficiency.

In a practical experimental realization, measurement outcomes may be
affected by the environment noise in the quantum channel or the dark
counts on the detector's side. We model this by adding the
white noise in the state preparation setup with
probability $p$. Then the effective state prepared is given by the
following equation

\begin{equation}
\begin{array}{lll}
\rho_{practical}=(1-p)\rho_{perfect}+\displaystyle  p\frac{I}{2},
\end{array}
\end{equation}
where $\rho_{perfect}$ is the perfect state preparation without
considering any noise. By considering the practical state
preparation $\rho_{practical}$, we get the following dimension
witness value

\begin{equation}
\begin{array}{lll}
T_{practical}=(1-p)\widetilde{T}_{j}.
\end{array}
\end{equation}
To violate the classical dimension witness upper bound $2$, the
background noise $p$ should satisfy $p<1-\displaystyle
\frac{2}{\widetilde{T}_{j}}$. If Bob has the perfect detection
efficiency in two bases ($\eta_0=\eta_1=1$), the maximal tolerated
background noise is $1-\displaystyle\frac{\sqrt{2}}{2}\simeq0.293$.
In the finite detection efficiency case, the corresponding maximal
tolerated background noise with different detection efficiency in
the symmetric and asymmetric case is given by Fig. 2 and Fig. 3
respectively.

\begin{figure}[!h]\center
\resizebox{9cm}{!}{
\includegraphics{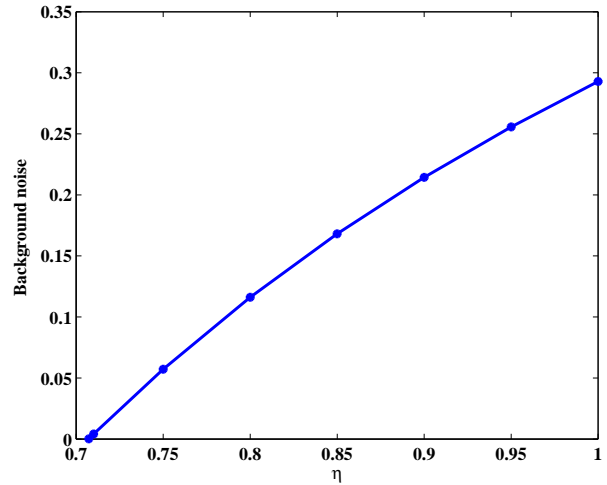}}
\caption{Maximal tolerable background noise as a function of the detection
efficiency in the symmetric case.}
\end{figure}

\begin{figure}[!h]\center
\resizebox{9cm}{!}{
\includegraphics{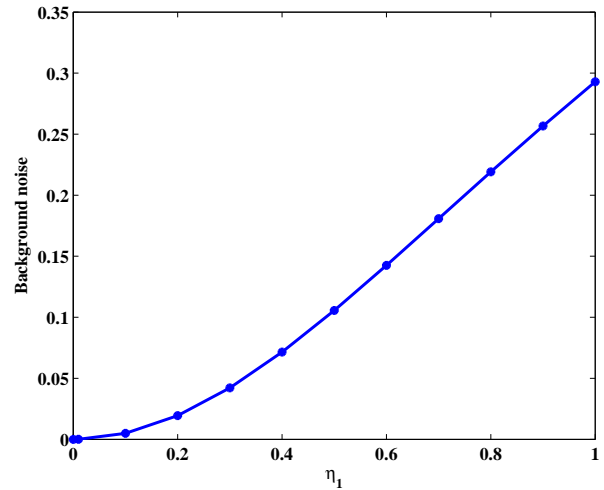}}
\caption{Maximal tolerable background noise as a function of the  detection
efficiency in the asymmetric case.}
\end{figure}

{\it Randomness generation certified with average guessing probability -}
In the randomness generation protocol we are given an infinite supply of pseudorandom numbers (PRN) which we assume to be independent of the devices that we are using. The aim of the protocol is to generate {\it certifiable } randomness (it is impossible in the case of pseudorandomness). We can use PRN in many different ways, effectively choosing the joint distribution of the inputs $a_0,a_1,y$. For example, if we discover that more randomness is generated for inputs $a_0=a_1=y=0$ then in a vast majority of rounds this input will be chosen. Other settings will be used only sporadically to estimate the value of $T$.

In the previous work \cite{li1}, we showed that the critical dimension
witness value to generate random number should be $2.64$ by using
the maximal guessing probability to estimate randomness (randomness
generation means Eve's maximal guessing probability should satisfy
$max_{b,a_0,a_1,y} p(b|a_0,a_1,y)<1$), which is obviously larger
than the classical dimension witness upper bound 2.

To generate much more randomness, we will apply new randomness
estimation methods, which should have two important properties. The
first property is that the randomness generation
efficiency should be larger than in the previous maximal guessing
probability method, the second is that the
random numbers should be certified as soon as the dimension witness value is larger
than the classical dimension witness upper bound. Now, we analyze
the randomness generation efficiency with partial average guessing
probabilities $p_{guess}^{(2)}$, $p_{guess}^{(3)}$ and full average
guessing probability $p_{guess}^{(4)}$ given by the following
equations

\begin{equation}
\begin{array}{lll}
p_{guess}^{(1)}=max_{b,a_0,a_1,y} p(b|a_0,a_1,y),\\
p_{guess}^{(2)}=\displaystyle \frac{1}{4}\sum_{a_1,y}max_b p(b|0,a_1,y),\\
p_{guess}^{(3)}=\displaystyle \frac{1}{4}\sum_{a_1,y}max_b p(b|1,a_1,y),\\
p_{guess}^{(4)}=\displaystyle \frac{1}{8}\sum_{a_0,a_1,y}max_b
p(b|a_0,a_1,y),
\end{array}
\end{equation}
where $p_{guess}^{(1)}$ is the maximal guessing probability, which
has been be applied to estimate the min-entropy function value
\cite{entropy} of the measurement outcomes in the previous work
\cite{li1}. Since $p_{guess}^{(1)}\geq max\{ p_{guess}^{(2)},
p_{guess}^{(3)},p_{guess}^{(4)}\}$ it is natural to think that the new methods will generate
much more randomness compared to the previous work.

In the experiment estimating the min-entropy on $p_{guess}^{(1)}$ or $p_{guess}^{(4)}$ corresponds to uniform distribution of the inputs, while using $p_{guess}^{(2)}$ implies choosing $a_0=0$ almost always and $p_{guess}^{(3)}$ almost never.

By considering different dimension witness value, we solve the
following optimization problem to estimate the guessing
probabilities $p_{guess}^{(i)}$

\begin{equation}
\begin{array}{lll}
maximize: p_{guess}^{(i)},\\
 subject~~to:  E_{a_0a_1y}=tr(\rho_{a_0,a_1}M_y^{b=0}),\\
~~~~~~~~~~~~~~\sum_{a_0,a_1,y}(-1)^{a_y}E_{a_0a_1y}=T,
\end{array}
\end{equation}
where $i=\{1,2,3,4\}$, $\rho_{a_0,a_1}=\displaystyle \frac{1}{2}
(I+\vec{s_{a_0a_1}}\cdot\vec{\sigma})$ and
$M_y^{b=0}=c_yI+\vec{t_y}\cdot\vec{\sigma}$ are arbitrary state
preparation and positive-operator valued measures (POVM) in the two
dimensional Hilbert space. The corresponding min-entropy function
system can be given by

\begin{equation}
\begin{array}{lll}
 H_{\infty}^{(i)}=-log_2p_{guess}^{(i)}.
\end{array}
\end{equation}
By using nonlinear optimization, the min-entropy function for
different dimension witness has been calculated and is given in Fig.
4. From the calculation result, we can find that the average
guessing probability methods can generate much more randomness
compared to the maximal guessing probability method, and random
numbers can be generated as soon as the dimension witness is larger
than the classical upper bound. For different dimension witness
value $T$, the optimal either $p_{guess}^{(2)}$ or $p_{guess}^{(4)}$
guessing probability should be chosen to generate randomness.
\begin{figure}[!h]\center
\resizebox{9cm}{!}{
\includegraphics{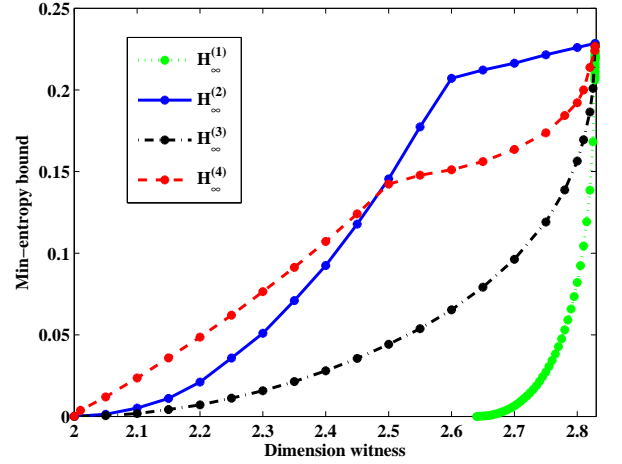}}
\caption{The min-entropy function
$H_{\infty}^{(1)}$,$H_{\infty}^{(2)}$,$H_{\infty}^{(3)}$ and
$H_{\infty}^{(4)}$ as a function of dimension witness value $T$. The
green dotted line is $H_{\infty}^{(1)}$, the blue solid line is
$H_{\infty}^{(2)}$, the black dot-dashed line is $H_{\infty}^{(3)}$,
and the red dashed line is $H_{\infty}^{(4)}$. To generate more
random numbers, the min-entropy function $H_{\infty}^{(4)}$ should be
applied to estimate the generated random number if the dimension
witness value satisfies $T<2.5$, while $H_{\infty}^{(2)}$ should be
used if $T\geq2.5$.}
\end{figure}

{\it Analytic bound on randomness -} We have numerically calculated the min-entropy function
based on different guessing probability methods in the previous
section. Now we find the upper bound of the guessing probability
$p_{guess}^{(2)}$, $p_{guess}^{(3)}$ and $p_{guess}^{(4)}$. Similar to the previous work
\cite{brunner2}, we assume Alice's and Bob's devices are governed by internal variables $\lambda$, and the distributions of these variables
is $q_{\lambda}$, where $\displaystyle\int  q_{\lambda}
d\lambda=1$. Since the observer has no access to the the value of the
variable $\lambda$, he will observe the following distribution in
practical experiment

\begin{equation}
\begin{array}{lll}
E_{a_0a_1y}=\displaystyle \int p(b=0|a_0,a_1,y,\lambda)q_{\lambda}
d\lambda.
\end{array}
\end{equation}

For a given internal parameter $\lambda$, the guessing
probabilities $p_{guess}^{(2)}$, $p_{guess}^{(3)}$ and
$p_{guess}^{(4)}$  change to

\begin{equation}
\begin{array}{lll}
{p_{guess}^{(2,\lambda)}}=\displaystyle \frac{1}{4}\sum_{a_1,y}max_b p(b|0,a_1,y,\lambda),\\
{p_{guess}^{(3,\lambda)}}=\displaystyle \frac{1}{4}\sum_{a_1,y}max_b p(b|1,a_1,y,\lambda),\\
{p_{guess}^{(4,\lambda)}}=\displaystyle
\frac{1}{8}\sum_{a_0,a_1,y}max_b p(b|a_0,a_1,y,\lambda).
\end{array}
\end{equation}

By considering the best guessing probability over Alice's different
inputs $a_0$ and $a_1$, an upper bound of the guessing probability can
be estimated by

\begin{equation}
\begin{array}{lll}
p_{guess}^{(j,\lambda)}\leq\displaystyle\frac{1}{2}max_{a_0,a_1}\sum_{y}max_b
p(b|a_0,a_1,y,\lambda)
\\~~~~~~~~\leq\displaystyle\frac{1}{2}+\displaystyle\frac{1}{2}cos(\frac{\theta_\lambda}{2})
\end{array}
\end{equation}
where $j=\{2,3,4\}$, $\theta_\lambda$ denotes the angle between
Bob's two measurements $M_0^{\lambda,b=0}$ and $M_1^{\lambda,b=0}$
by considering the best guessing strategy, more detailed analysis
was given in Ref. \cite{brunner2}. Since the correlation between the
guessing probability and parameter $\theta_\lambda$ has been
established, we will prove the relationship between the dimension
witness value $T^{\lambda}$ and the parameter $\theta_\lambda$

\begin{equation}
\begin{array}{lll}
T^{\lambda}\\
=E_{000}^{\lambda}+E_{001}^{\lambda}+E_{010}^{\lambda}-E_{011}^{\lambda}-E_{100}^{\lambda}+E_{101}^{\lambda}
-E_{110}^{\lambda}-E_{111}^{\lambda}\\
=tr[(\rho_{00}^{\lambda}-\rho_{10}^{\lambda})M_0^{\lambda,b=0}]+tr[(\rho_{00}^{\lambda}-\rho_{01}^{\lambda})M_1^{\lambda,b=0}]\\
~~+tr[(\rho_{01}^{\lambda}-\rho_{11}^{\lambda})M_0^{\lambda,b=0}]+tr[(\rho_{10}^{\lambda}-\rho_{11}^{\lambda})M_1^{\lambda,b=0}]\\
=\displaystyle\frac{1}{2}[(\vec{s^{\lambda}_0}-\vec{s^{\lambda}_2})\cdot\vec{t^{\lambda}_0}+(\vec{s^{\lambda}_0}-\vec{s^{\lambda}_1})\cdot\vec{t^{\lambda}_1}\\~~~~~~
+(\vec{s^{\lambda}_1}-\vec{s^{\lambda}_3})\cdot\vec{t^{\lambda}_0}+(\vec{s^{\lambda}_2}-\vec{s^{\lambda}_3})\cdot\vec{t^{\lambda}_1}
]\\
=\displaystyle\frac{1}{2}[(\vec{s^{\lambda}_0}-\vec{s^{\lambda}_3})\cdot(\vec{t^{\lambda}_0}+\vec{t^{\lambda}_1})+
(\vec{s^{\lambda}_1}-\vec{s^{\lambda}_2})\cdot(\vec{t^{\lambda}_0}-\vec{t^{\lambda}_1})
]\\
\leq|\vec{t^{\lambda}_0}+\vec{t^{\lambda}_1}|+|\vec{t^{\lambda}_0}-\vec{t^{\lambda}_1}|\\
\leq\sqrt{2+2cos(\theta_\lambda)}+\sqrt{2-2cos(\theta_\lambda)},
\end{array}
\end{equation}
where $\vec{s^{\lambda}_0}\equiv\vec{s^{\lambda}_{00}}$,
$\vec{s^{\lambda}_1}\equiv\vec{s^{\lambda}_{01}}$,
$\vec{s^{\lambda}_2}\equiv\vec{s^{\lambda}_{10}}$,
$\vec{s^{\lambda}_3}\equiv\vec{s^{\lambda}_{11}}$. The first
inequality uses $|\vec{s^{\lambda}_0}-\vec{s^{\lambda}_3}|\leq2$ and
$|\vec{s^{\lambda}_1}-\vec{s^{\lambda}_2}|\leq2$, the second
inequality can be proved by considering
$|\vec{t^{\lambda}_0}+\vec{t^{\lambda}_1}|+|\vec{t^{\lambda}_0}-\vec{t^{\lambda}_1}|=\sqrt{|\vec{t^{\lambda}_0}|^2+|\vec{t^{\lambda}_1}|^2+2|\vec{t^{\lambda}_0}||\vec{t^{\lambda}_1}|cos(\theta_\lambda)}+
\sqrt{|\vec{t^{\lambda}_0}|^2+|\vec{t^{\lambda}_1}|^2-2|\vec{t^{\lambda}_0}||\vec{t^{\lambda}_1}|cos(\theta_\lambda)}$
reach the maximum  if
$|\vec{t^{\lambda}_0}|=|\vec{t^{\lambda}_1}|=1$. Based on the given
internal variables $\lambda$, we get the relationship between the
guessing probability ${p_{guess}^{(j,\lambda)}}$ and the dimension
witness value $T^{\lambda}$ as the following inequality

\begin{equation}
\begin{array}{lll}
 p_{guess}^{(j,\lambda)}\leq\displaystyle
\frac{1}{2}+\displaystyle\frac{1}{2}\sqrt{\frac{1+\sqrt{1-(\frac{(T^{\lambda})^2-4}{4})^2}}{2}}
\\
~~~~~~~~\equiv f(T^{\lambda}),
\end{array}
\end{equation}
where the analysis uses the inequality
$\cos(\theta_\lambda)\leq\sqrt{1-(\frac{(T^{\lambda})^2-4}{4})^2}$.
Note that function $f(T^{\lambda})$ is concave and decreasing, we
will apply this property to prove the relationship between the
practical experimental estimated value $T$ and $p_{guess}^{(j)}$.

Since the internal variables $\lambda$ can not be detected in practical
experiment, we can only get the observed dimension witness value $T$
as

\begin{equation}
\begin{array}{lll}
T=\displaystyle \int T^{\lambda} q_{\lambda}  d\lambda.
\end{array}
\end{equation}
Based on the observed dimension witness value $T$, upper bound of
the guessing probability $p_{guess}^{(j)}$ is

\begin{equation}
\begin{array}{lll}
p_{guess}^{(j)}=\displaystyle \int p_{guess}^{(j,\lambda)}
q_{\lambda}
d\lambda\\
~~~~~~~~\leq\displaystyle \int f(T^{\lambda}) q_{\lambda}
d\lambda\\
~~~~~~~~\leq f(\displaystyle \int T^{\lambda} q_{\lambda} d\lambda)\\
~~~~~~~~=f(T)\\
~~~~~~~~=\displaystyle\frac{1}{2}+\displaystyle\frac{1}{2}\sqrt{\frac{1+\sqrt{1-(\frac{T^2-4}{4})^2}}{2}},
\end{array}
\end{equation}
where the first inequality uses the previous result, the second one
applies Jensen's inequality and concavity property of $f$. By using
this bound on the guessing probability, we calculate the min-entropy
function $-log_2(f(T))$ with different dimension witness $T$ in Fig.
5. From the calculation result, we can find that the maximal
min-entropy function is 0.228 when the dimension witness reaches
$2\sqrt{2}$, while the min-entropy function is larger than $0$ if
the dimension witness is larger than the classical dimension witness
upper bound.

\begin{figure}[!h]\center
\resizebox{9cm}{!}{
\includegraphics{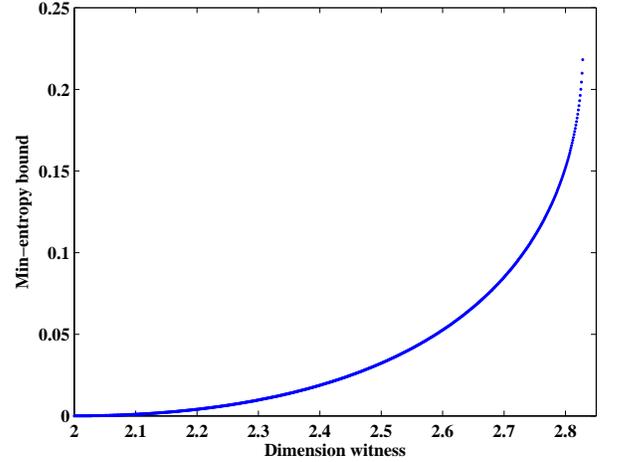}}
\caption{Min-entropy function $-log_2(f(T))$ with different
dimension witness $T$, the figure is based on the analytic result,
which is the lower bound on the previous numerical calculation
result $H_{\infty}^{(2)}$, $H_{\infty}^{(3)}$ and
$H_{\infty}^{(4)}$.}
\end{figure}

{\it Conclusion -}We have calculated the critical detection efficiency for semi-device independent random number generation in
the symmetric and asymmetric case. The maximal tolerable white noise has also been
analyzed. To improve the randomness generation,
three type of averaging guessing probability have been tested
in our work. We also give the general analytical
relationship between the average guessing probability and the
dimension witness. Our analysis result can be directly applied in
practical experimental realization and the future research on other
semi-device independent quantum information processing protocols.

 To further decrease the critical
detection efficiency and improve the random number generation
efficiency in the SDI protocols is an open problem for the future
research.

{\it Acknowledgements}  The author Hong-Wei Li thanks Yao Yao for
his helpful discussion.  The authors are supported by the the
National Natural Science Foundation of China (Grant Nos. 11304397,
61201239, 61205118 and 61475148), China Postdoctoral Science
Foundation (Grant No. 2013M540514) Anhui Provincial Natural
Science Foundation(Grant No. 1408085QF102), the ERC grant QOLAPS, the FNP grant TEAM and the NCN grant 2013/08/M/ST2/00626.
$^a$yinzheqi@mail.ustc.edu.cn, $^b$dokmpa@univ.gda.pl,
$^c$zfhan@ustc.edu.cn.

\end{document}